\documentclass[sigconf]{acmart}
\renewcommand\footnotetextcopyrightpermission[1]{}

\AtBeginDocument{%
  }

\thanks{This is a preprint accepted for ICSE 2026 - Demonstration Track.}

\usepackage[acronym]{glossaries}
\usepackage{caption}
\usepackage{tcolorbox}
\usepackage{multirow}
\usepackage{stfloats}
\usepackage{tikz}
\usepackage{tabularx}
\usepackage{xcolor} 

\definecolor{myblue}{HTML}{002C4A}

\newcommand*\bluecircled[1]{%
  \tikz[baseline=(char.base)]{
    \node[
      shape=circle,
      draw=myblue ,   
      fill=myblue,   
      text=white,   
      thick,
      inner sep=0.2pt
    ] (char) {#1};}}
    
\tcbuselibrary{breakable}

\begin{document}
\newacronym{OAS}{OAS}{OpenAPI Specification}
\newacronym{SUT}{SUT}{Service Under Test}
\title{RESTifAI: LLM-Based Workflow for Reusable REST API Testing}

\author{Leon Kogler}
\email{leon.kogler@casablanca.at}
\orcid{0009-0000-7319-2398}

\author{Maximilian Ehrhart}
\email{maximilian.ehrhart@casablanca.at}
\orcid{0000-0002-9554-0231}
\affiliation{%
  \institution{CASABLANCA hotelsoftware GmbH}
  \city{Schönwies}
  \country{Austria}
}

\author{Benedikt Dornauer}
\orcid{0000-0002-7713-4686}
\email{benedikt.dornauer@uibk.ac.at}

\affiliation{
  \institution{University of Innsbruck}
  \city{Innsbruck}
  \country{Austria}
}

\author{Eduard Paul Enoiu}
\orcid{0000-0003-2416-4205}
\email{eduard.paul.enoiu@mdh.se}

\affiliation{
  \institution{Mälardalen University}
  \city{Mälardalen}
  \country{Sweden}
}

\newtcolorbox{casablancabox}[1][]{%
  colback=gray!10,
  colframe=black!50,
  boxrule=0.1pt,
  arc=1pt,
  lefttitle=0.15cm,
  righttitle=0.15cm,
  leftupper=0.15cm,
  rightupper=0.15cm,
  title=Insights from CASABLANCA,
  breakable,
  #1 
}

\begin{abstract} 
With this paper, we introduce RESTifAI, an LLM-driven approach for generating reusable, CI/CD-ready REST API tests, following the happy-path approach. 
Unlike existing tools that often focus primarily on internal server errors, RESTifAI systematically constructs valid test scenarios (happy paths) and derives negative cases to verify both intended functionality (2xx responses) and robustness against invalid inputs or business-rule violations (4xx responses). 
The results indicate that RESTifAI performs on par with the latest LLM tools, i.e., AutoRestTest and LogiAgent, while addressing limitations related to reusability, oracle complexity, and integration. To support this, we provide common comparative results and demonstrate the tool’s applicability in industrial services. For tool demonstration, please refer to \url{https://www.youtube.com/watch?v=2vtQo0T0Lo4}. RESTifAI is publicly available at \url{https://github.com/casablancahotelsoftware/RESTifAI}.
\end{abstract}

\ccsdesc[300]{Software and its engineering~API languages}
\ccsdesc[500]{Software and its engineering~Acceptance testing}
\ccsdesc[300]{Software and its engineering~Software testing and debugging}

\keywords{REST API Testing, OpenAPI, Automated Test Generation, Large Language Model}

\pagestyle{plain}
\maketitle

\section{Introduction}
Migrating from monolithic architectures to microservices, alongside the transition from native software to web-based solutions, has intensified reliance on REST APIs \cite{lamotheSystematicReviewAPI2022}. This trend is further motivated by the growing reliance on external services that expose APIs. Many services now follow the \textit{\gls{OAS}} standard for consistency. Although this standard effectively documents service capabilities and simplifies integration, it does not guarantee correctness. Hence, problems may emerge during integration due to incorrect API behaviors, or, more critically, during operation when unannounced background modifications, such as updates to code or infrastructure, trigger unforeseen failures and disruptions. Thus, these risks emphasize the importance of systematically and continuously testing REST APIs ~\cite{kotsteinWhichRESTfulAPI2021, golmohammadiTestingRESTfulAPIs2024a}. In this way, several tools were developed, which were reviewed (till 2023) by \citet{golmohammadiTestingRESTfulAPIs2024a}, showing that fuzzing is the predominant technique, besides rule-based, property-based, and search-based approaches. More recent tools incorporate Large Language Models (LLMs), demonstrating new capabilities and improved performance. In particular, \textit{AutoRestTest}\cite{stennettAutoRestTestToolAutomated2025a} (or a derivative:\textit{LlamaRestTest} \cite{kimLlamaRestTestEffectiveREST2025} leveraging Llama models), \textit{LogiAgent} \cite{zhangLogiAgentAutomatedLogical2025}, and \textit{APITestGenie} \cite{pereiraAPITestGenieAutomatedAPI2024}. 

Before initiating the development of our tool, we first examined the limitations of existing approaches. One of the practical limitations is that existing tools are often not designed to generate test suites that can be systematically re-executed. For example, AutoRestTest incorporates mutators that create a large number of request variations, making the approach less suitable for repeated testing. To the best of our knowledge, APITestGenie is the only other LLM-based tool explicitly targeting reusable, CI/CD-ready test suites. However, since it requires additional business requirements in natural language, we excluded it from our evaluation. Apart from that, APITestGenie`s authors \cite{pereiraAPITestGenieAutomatedAPI2024} note that only 57.3\% of the APITestGenie-generated cases execute successfully, due to hallucinations, setup issues, and misinterpretation of \gls{OAS}. Another limitation of tests generated by tools is the lack of support for incorporating API-specific information, such as authentication credentials or API keys. Moreover, tests generated by most current API testing tools are often limited by the capabilities of their test oracles. Tools, such as the often-used EvoMaster \cite{arcuriEvoMasterEvolutionaryMulticontext2018} or AutoRestTest, primarily adopt random and mutation-based parameter generation strategies and rely on automated $500$ response code oracles \cite{kimAutomatedTestGeneration2022, golmohammadiTestingRESTfulAPIs2024a}. While this strategy appears to be effective in identifying server crashes \cite{stennettAutoRestTestToolAutomated2025a, kimAutomatedTestGeneration2022}, it might be insufficient for validating realistic scenarios and intended functional behavior. Such tools do not check that valid inputs yield successful (\textit{2xx}) responses or that invalid inputs are correctly rejected with client-error (\textit{4xx}) responses. 

To address these limitations, we present RESTifAI, a LLM-based workflow whose novelty derives from automatically generating positive tests (happy paths), which confirm correct system behavior under valid inputs, and systematically deriving negative tests from these happy paths, that validate robustness under invalid or unexpected conditions. 

\section{Approach and Architecture}

\begin{figure*}[t]
  \centering
  \includegraphics[width=1 \linewidth]{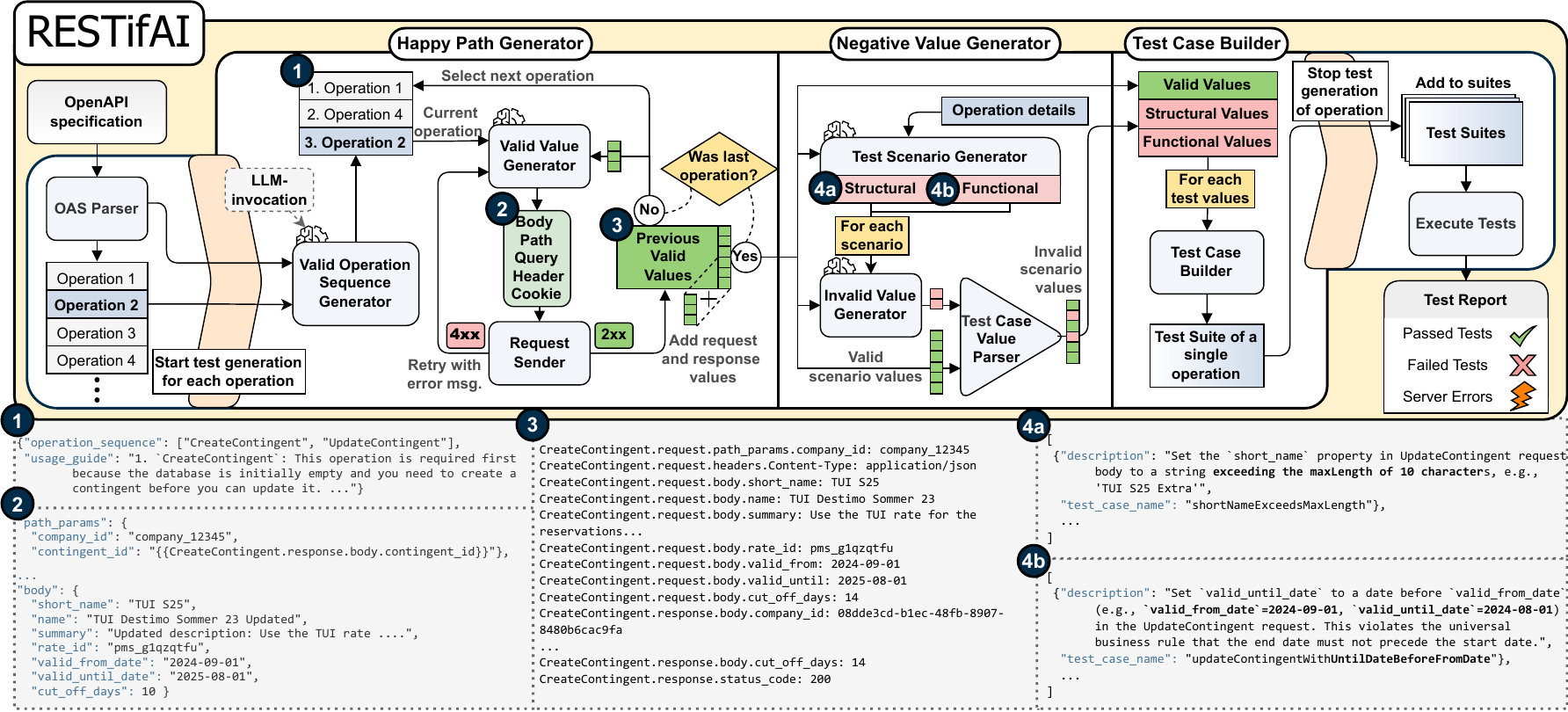}
  \caption{RESTifAI workflow with illustrative outputs from four intermediate steps. \protect\bluecircled{1}, \protect\bluecircled{2}, \protect\bluecircled{4} are structured LLM outputs.}
  \Description{Overview of RESTifAI. The framework parses the OpenAPI specification and generates test cases through three stages: (1) Happy Path Generator for valid request sequences, (2) Negative Value Generator for structural and functional fault scenarios, and (3) Test Case Builder for assembling and executing test suites, producing detailed test reports.}
  \label{fig:overview}
\end{figure*}

Unlike fully autonomous agentic systems such as LogiAgent, RESTifAI adopts a workflow-based approach, using LLMs selectively to ensure more reliable test generation \cite{zhangBuildingEffectiveAI, khotDecomposedPromptingModular2022}. Overall, it can be seen as exploratory testing, yet also as requirements-based, since it relies on the \textit{\gls{OAS}} specification. The main components of RESTifAI include a \textit{Happy Path Generator}, a \textit{Negative Value Generator}, and a \textit{Test Case Builder}, as illustrated in Figure \ref{fig:overview}. As input, our tool takes the \textit{\gls{OAS}} of the \textit{\gls{SUT}} and, as a result, generates a test suite (logically structured test cases) for each operation, executes the tests, and returns a test report including passed tests, failed tests, and highlighted server errors. The steps are detailed further: 

\paragraph{Happy Path Generation}
The \textbf{Happy Path Generator} aims to create a valid scenario that results in a 2xx response for the operation under test, while tracking parameters and values of request and response properties, as well as their dependencies. Therefore, the Happy Path Generator utilizes an LLM to generate a sequence of dependent operations required to produce a valid request for the final operation (the operation under test). The LLM also provides a natural-language usage guide outlining these dependencies and guiding the \textbf{Value Generator} on how parameter values should be derived \bluecircled{1}. Next, in an iterative process, RESTifAI attempts to generate valid values \bluecircled{2} for each operation in the sequence, based on the operation’s specification, the previously generated usage guide, and the execution trace of previously successfully executed dependent operations \bluecircled{3}. This execution trace is represented by key-value pairs, where the keys identify the exact location of values within the test scenario. Values may be either \texttt{GENERATED}, representing newly created inputs, or \texttt{DEPENDENT}, referencing keys of previously stored values. Starting with an initial empty execution trace, the \textbf{Value Generator} generates \texttt{Body, Path, Query, Header,} and \texttt{Cookie} parameters for the first operation in the sequence. The \textbf{Request Sender} then uses the generated values to send a request to the service endpoint. If the request yields a 4xx response, the Value Generator retries, incorporating the error message into the regeneration process. Upon receiving a successful 2xx response, the request and response values are parsed into key–value pairs and stored in the execution trace. This procedure repeats for each operation in the sequence until the final operation is executed successfully, at which point RESTifAI proceeds to generate negative values.

\paragraph{Negative Value Generator}
Using the final “Happy Path” data and the specification of the operation under test, the \textbf{Test Scenario Generator} employs an LLM to produce oracles that define invalid parameter values expected to result in 4xx client error responses. These are split into two categories of oracles: \textit{Structural oracles} \bluecircled{4a} test the \textit{\gls{SUT}} against invalid inputs derived from constraints in the \textit{\gls{OAS}}, such as incorrect data types or formats, while \textit{Functional oracles} \bluecircled{4b}, in contrast, verify business logic constraints not captured in the \textit{\gls{OAS}} by leveraging the LLM’s domain-specific knowledge to define logical limitations of the \textit{\gls{SUT}}. For each generated test scenario, the \textbf{Invalid Value Generator} invokes the LLM to generate invalid parameter values using the parameter keys recorded in the valid execution trace from the \textbf{Happy Path Generator}. The \textbf{Test Case Value Parser} then replaces the valid values in the original scenario with the corresponding invalid values. This ensures that the negative test case triggers a 4xx response due to the substituted values while all other parameters remain valid.

\paragraph{Test Case Builder}
Based on the positive scenario values generated by the \textbf{Happy Path Generator} and the multiple negative scenario values generated by the \textbf{Negative Value Generator}, the \textbf{Test Case Builder} generates an executable Postman collection for each of them. By maintaining the scenario values in a standardized key–value format the testing framework can be chosen independently, requiring modification only to the \textbf{Test Case Builder} to adapt to different test frameworks.

To ensure reproducibility and isolation RESTifAI allows the integration of environment initialization scripts in various formats. Such scripts are executed before generating a new Happy Path and before executing each individual test case, thereby maintaining the \textit{\gls{SUT}} in a controlled and consistent test environment. For instance, an initialization script may reset the database to a predefined state, ensuring test repeatability. Additionally, RESTifAI can be guided with human-provided natural language during Happy Path generation by adding information not specified in the \textit{\gls{OAS}}, such as authentication credentials or pre-configured database identifiers, to the LLM’s prompt.

RESTifAI includes a command-line interface (CLI) that allows users to customize the test generation process by selecting the type of test cases, specifying environment initialization scripts, or providing user input. For enhanced usability and a more interactive experience, a Python-based graphical frontend is provided, allowing users to efficiently configure test scenarios, execute them, and visually inspect the resulting test reports.

\section{Results and Comparison to Current Tools}
To assess the quality of RESTifAI in automatic REST-API testing, we conducted a comparison with AutoRestTest and LogiAgent, all leveraging the same Azure OpenAI GPT-4.1-mini LLM\footnote{The replication package, along with a detailed execution description, is available on \url{https://github.com/casablancahotelsoftware/RESTifAI}.}.

\begin{table}[H]
    \caption{Number of Operations Covered (\#OC) and the Server Errors (\#SE) detected by the tools on the services.}
    \label{tab:oc-se-coverage}
    \centering
    \begin{tabular}{lcccccc}\toprule
        \multirow{2}{*}{Service} &  \multicolumn{2}{c}{AutoRestTest}&\multicolumn{2}{c}{LogiAgent}& \multicolumn{2}{c}{RESTifAI}\\
        \cmidrule(lr){2-3} \cmidrule(lr){4-5} \cmidrule(lr){6-7}
        &  \#OC&\#SE&  \#OC&\#SE& \#OC&\#SE\\\midrule
        FDIC&  6&\textbf{41}&6&0& 6&0\\
        Genome Nexus&  23&0&19&0& 23&0\\
        Languagetool&  2&0&1&0& 2&\textbf{1}\\
        OhSome&  33&0& 4&3& \textbf{128}&\textbf{50}\\
        Restcountries&  \textbf{22}&83&13&0& 20&0\\
        \bottomrule
        \end{tabular}
\end{table}

The evaluation was performed on a set of standard benchmarks such as open-source services (Languagetool, Genome Nexus, and Restcountries), which we run locally to enable code-coverage analysis using the JaCoCo library, and online services (OhSome and FDIC), for which source code is inaccessible and code-coverage metrics cannot be obtained. References to the services can be found in our Replication Package. OhSome was explicitly included because it is considered one of the most complex services in the AutoRestTest study, and AutoRestTest was the only tool to generate 12 successful requests for it.

Due to conceptual differences between the tools, we standardized test generation using execution time to ensure a fairer comparison. Our tool stops after a set number of test suites, matching the number of API operations, while tools like LogiAgent and AutoRestTest keep generating test cases indefinitely, potentially leading to longer runtimes under the same conditions. The execution time of RESTifAI is then applied to AutoRestTest and LogiAgent for each service (see replication package for execution time).

\textbf{Operation Coverage}
 is a key metric for evaluating automated API testing tools \cite{stennettAutoRestTestToolAutomated2025a, golmohammadiTestingRESTfulAPIs2024a, kimAutomatedTestGeneration2022} and plays a fundamental role for RESTifAI, as test cases are generated only if a valid request for the operation under test can be produced. While AutoRestTest optimizes exploration of successful 2xx responses, RESTifAI achieved 128 out of 134 successful operations on the OhSome service, compared to 33 operations covered by AutoRestTest, as shown in Table \ref{tab:oc-se-coverage}. These results highlight the effectiveness of targeted value generation over mutation-based approaches, particularly as the number of operation parameters increases.
 
The number of detected \textbf{Server Errors} is commonly used to compare fuzzing-based testing approaches \cite{golmohammadiTestingRESTfulAPIs2024a}. While AutoRestTest is primarily optimized for uncovering Server Errors, RESTifAI is designed to generate realistic input data that effectively tests both 2xx success responses and 4xx client errors. As shown for OhSome in Table \ref{tab:oc-se-coverage}, RESTifAI also performs well in uncovering Server Errors, especially in scenarios requiring the precise generation of real-world edge cases. 

\begin{table}[h]
    \caption{Comparison of BC (Branch-Coverage), LC (Line-Coverage), and MC (Method-Coverage) for local services.}
    \label{tab:coverage}
    
    \centering
    \begin{tabular}{clccc}\toprule
         &  Service & AutoRestTest&  LogiAgent&  RESTifAI\\\midrule
 \multirow{3}{*}{BC}
 & Genome Nexus& 0\%& 0\%&0\%\\
 & Languagetool& 16\%& 7\%&19\%\\
 & Restcountries& 88\%& 40\%&74\%\\ \midrule
  \multirow{3}{*}{LC}
 & Genome Nexus
& 65\%& 60\%&65\%\\
& Languagetool
& 27\%& 16\%&28\%\\
 & Restcountries& 78\%& 49\%&72\%\\ \midrule
  \multirow{3}{*}{MC}
 & Genome Nexus
& 45\%& 39\%&45\%\\
& Languagetool
& 28\%& 16\%&30\%\\
 & Restcountries& 85\%& 68\%&83\%\\
 \bottomrule
    \end{tabular}

\end{table}

\begin{table}[h]
    \caption{Number of Test Cases (\# TC) generated and the number of Tokens per Test Case (Tokens/TC) }
    \label{tab:test-cases}
    \centering
    \begin{tabular}{lcccc}\toprule
         \multirow{2}{*}{Service} &  \multicolumn{2}{c}{LogiAgent}&  \multicolumn{2}{c}{RESTifAI}\\
         \cmidrule(lr){2-3} \cmidrule(lr){4-5}
         & \#TC&Tokens/TC& \#TC&Tokens/TC\\\midrule
         FDIC& 26&37001& 133&88762\\
         Genome Nexus& 38&62071& 326&22258\\
         Languagetool& 6&46689& 35&3585\\
         OhSome&  234&48036& 2213&37947\\
         Restcountries& 27&62143& 191&9296\\ \midrule
         Average& & 51188& &32370\\
         \bottomrule
    \end{tabular}
\end{table}

We match AutoRestTest's performance in line, branch, and method coverage across various services (see Table \ref{tab:coverage}). In general, it can be seen that higher OC tends to lead to higher \textbf{Code Coverage}.

RESTifAI creates individual oracles for each new test case similar to LogiAgent, which is not possible with AutoRestTest. The workflow-based approach combined with selective LLM utilization results in a significantly larger \textit{Number of Test Cases}, while using significantly fewer  \textit{Tokens per Test Case}, as shown in Table~\ref{tab:test-cases}. We required more tokens on FDIC because of storing the large payloads returned from the service in our key-value store. 

\begin{casablancabox}[title=Application to a Real-World Use Case]

Beyond our benchmarks on well-established API services, RESTifAI was applied to two microservices from CASABLANCA hotelsoftware: the \textit{Guest-Review Service} and the \textit{Inventory Service}. Real-world services often rely on external systems or require specific parameters like a fixed 'company\_id', so automated testing tools must handle these constraints. RESTifAI allows users to provide such parameters or define a custom sequence of operations via the CLI to generate a successful happy path, thereby offering a clear advantage over existing tools.

\vspace{0.2cm}
\begin{center}
\captionof{table}{Classification of reusable test cases}
\begin{tabular}{ccccl}
\toprule
 & Bug & Enhancement & Invalid & Passed \\
\midrule
Guest-Review & 2 & 12 & 9  & 51 \\
Inventory    & 11& 6& 30& 91 \\
\bottomrule
\end{tabular}
\label{tab:casablanca}
\end{center}
\vspace{0.2cm}

In Table \ref{tab:casablanca}, we classified the failed generated test cases with domain-experts into Bug, Enhancement and Invalid test cases following the metrics defined in \cite{zhangLogiAgentAutomatedLogical2025}. The high number of invalid test cases in the Inventory Service can be attributed to the fact that certain business rules are not explicitly captured in the API contract, leading to invalid test cases. In the Guest-Review Service, 60.87\% of failed test cases revealed real bugs or enhancements. 

For our services, we identified, e.g. the following circumstances:  

\textbf{Case 1: Structural Error}  
The test case 'roomtypeIdWrongType\_ST' revealed a structural inconsistency where an integer was provided for the 'room\_type\_id' parameter instead of the expected string.

\textbf{Case 2: Logical Error}  
In the Inventory Service, our tool identified an issue with date-range constraints: the API allowed the 'until' date to be set before the 'from' date, which violates the expected business logic; see Fig. \ref{fig:overview}. 

Both cases highlight a key advantage of RESTifAI over approaches that rely solely on automated 5xx response code oracles. In each case, the services erroneously returned a 200 response code instead of the appropriate 400 Bad Request response code. Tools such as AutoRestTest, which detect only 5xx response codes, are not able to identify such issues.

\end{casablancabox}

\section{Discussion and Conclusion}

RESTifAI advances LLM-based API testing by coupling happy-path generation with systematic negative derivation. Thereby, the tool complements fuzzers by prioritizing correctness (2xx/4xx responses), emphasizing OAS-conformant behavior and business-rule validation, whereas fuzzers excel at broad input exploration and crash detection. The results indicate that we compete with the latest REST API test generation tools. Additionally, compared to existing approaches, RESTifAI produces fully automated and re-executable CI/CD-ready tests, while also offering simple integration benefits, as demonstrated in the use case or demo video. A current limitation of RESTifAI is the absence of assertions that validate the functional behavior of the underlying business logic, a recurring shortcoming in existing tools, which we aim to address in future work. 

As RESTifAI is designed to generate more complex oracles, there is currently no universal method to automatically measure their correctness. At present, domain experts must verify the generated tests, which introduces observer bias and complicates validity comparisons across tools, especially when \textit{\gls{SUT}} requirements are rarely defined. One promising direction may be to employ LLMs as judges to mitigate observer bias. Currently, RESTifAI reuses the same operation sequences for both happy-path and negative test generation. While this aims to improve validity, it also restricts the exploration of diverse scenarios. This limitation could be mitigated through more dynamic strategies, such as multi-agent or human-in-the-loop approaches. 

\begin{acks}
This work was supported by and done within the scope of the ITEA4 GENIUS project, which was nationally funded by FFG with grant 921454. Thanks for the support from Diffblue, Peter Schrammel. Grammarly and ChatGPT were used to enhance text quality.

\end{acks}
\balance
\bibliographystyle{ACM-Reference-Format}
\bibliography{references}

\end{document}